\documentclass[a4paper,11pt]{article}
\usepackage{pos}

\title{\textbf{The $\gamma Z$-box correction and its impact on parity-violating deep-inelastic scattering}
}

\author*[a,b]{Balma Duch}
\author[a,b]{Pere Masjuan}
\author[c]{Hubert Spiesberger}

\affiliation[a]{Grup de Física Teòrica, Departament de Física, Universitat Autònoma de Barcelona (UAB), Campus UAB, E-08193 Bellaterra (Barcelona), Spain}

\affiliation[b]{Institut de Física d'Altes Energies (IFAE) and the Barcelona Institute of Science and Technology (BIST), Campus UAB, E-08193 Bellaterra (Barcelona), Spain}

\affiliation[c]{PRISMA+ Cluster of Excellence, Institut für physik, Johannes Gutenberg-Universität, 55099 Mainz, Germany}

\emailAdd{bduch@ifae.es}
\emailAdd{masjuan@ifae.es}
\emailAdd{spiesber@uni-mainz.de}

\abstract{The $\gamma Z$-box correction is an important electroweak contribution to precision parity-violation measurements, but its conventional low-energy treatment relies on an ambiguous effective quark-mass prescription. We revisit this correction using a finite-mass calculation that provides a well-defined perturbative contribution without introducing such an auxiliary scale. Applying the result to the Jefferson Lab PVDIS measurement shifts the extracted electron-quark couplings at the $10^{-3}$ level. We discuss the implications of this shift for precision electroweak tests and the need for a consistent separation of perturbative and non-perturbative hadronic contributions.
}

\FullConference{The 33rd International Workshop on Deep Inelastic Scattering and Related Subjects (DIS2026)\\
4 - 8 May 2026\\
Bologna, Italy\\}

\begin{document}
\maketitle

\section{Introduction}

Precision parity-violating measurements provide sensitive tests of the electroweak interaction at energies well below the electroweak scale. Their interpretation can be formulated in terms of effective electron-quark couplings, whose relation to the Standard Model parameters requires the inclusion of electroweak radiative corrections. Among these corrections, the $\gamma Z$-box contribution is particularly relevant for parity-violating electron scattering and has important implications for the extraction of low-energy electroweak couplings.

The treatment of the low-energy $\gamma Z$-box has a long history. The calculation used in early analyses of atomic parity violation was developed by Marciano and Sirlin \cite{Marciano:1982mm}, following the prescription introduced by Marciano and Sanda \cite{Marciano:1978ed}. In that approach, the external momenta are set to zero and the infrared singularity associated with the soft photon is regulated by introducing a finite quark mass, interpreted as the scale below which quarks behave approximately as free particles. This prescription was subsequently adopted in low-energy electroweak analyses \cite{Erler:2013xha}. However, the use of an effective quark mass as an infrared regulator is not uniquely defined, and different treatments have led to different results for the $\gamma Z$-box contribution.

In our recent work \cite{Duch:2026ujd}, we revisited the one-loop $\gamma Z$-box contribution to electron-quark scattering by retaining finite electron and quark masses throughout the calculation. This allows for a direct treatment of the infrared behaviour of the amplitude and allows the low-energy limit to be studied without introducing an effective constituent-quark mass as an additional prescription. The calculation also determines the constant terms accompanying the logarithmic contribution directly from the amplitude.

In this proceedings contribution, we investigate the phenomenological relevance of our result for parity-violating deep-inelastic scattering. We revisit the original analysis of the Jefferson Lab electron-deuteron measurement \cite{Wang:2014guo}, using our corrected $\gamma Z$-box contribution in the $Q^2\rightarrow0$ limit to determine the resulting shift in the extracted electron-quark couplings. We finally discuss the relation between the perturbative contribution obtained from our calculation and the remaining non-perturbative hadronic correction.

\section{The $\gamma Z$-box at low momentum transfer}
\label{sec:gammaZbox}
We summarize here the part of our one-loop calculation that is relevant for the phenomenological analysis below. In contrast to the conventional low-energy treatment, we retain finite electron and quark masses, $m_e$ and $m_q$, throughout the calculation and take the low-energy limit only after evaluating the full amplitude. At zero momentum transfer, the box contribution is finite, while the threshold limit develops a Coulomb singularity associated with the interaction of the external charged particles. This singularity has the form of a Sommerfeld enhancement and is velocity dependent, and therefore does not contribute to the local low-energy couplings. Following Ref.~\cite{Duch:2026ujd}, we subtract this contribution and retain the remaining finite terms.

The expressions obtained after taking one of the fermion masses to zero are particularly simple. For vanishing quark mass, $m_q\to0$, we find
\begin{align}
\delta_{\rm box}C_{1q}
&=
-\frac{\alpha}{4\pi}Q_eQ_q6g_{VA}^{eq}
\left[
\ln\left(\frac{M_Z^2}{m_e^2}\right)+\frac{5}{6}
\right],
\label{eq:proceedings_C1}\\
\delta_{\rm box}C_{2q}
&=
-\frac{\alpha}{4\pi}Q_eQ_q\,6g_{AV}^{eq}
\left[
\ln\left(\frac{M_Z^2}{m_e^2}\right)+\frac{3}{2}
\right].
\label{eq:proceedings_C2}
\end{align}
Conversely, taking $m_e\to0$ while retaining a finite quark mass gives
\begin{align}
\delta_{\rm box}C_{1q}
&=
-\frac{\alpha}{4\pi}Q_eQ_q\,6g_{VA}^{eq}
\left[
\ln\left(\frac{M_Z^2}{m_q^2}\right)+\frac{3}{2}
\right],
\label{eq:proceedings_C1_me0}
\\
\delta_{\rm box}C_{2q}
&=
-\frac{\alpha}{4\pi}Q_eQ_q\,6g_{AV}^{eq}
\left[
\ln\left(\frac{M_Z^2}{m_q^2}\right)+\frac{5}{6}
\right].
\label{eq:proceedings_C2_me0}
\end{align}
Thus, the large logarithm is present in both limits, but both its scale and the accompanying constant term depend on which fermion mass is retained. This differs from the results adopted in previous low-energy analyses~\cite{Marciano:1982mm,Erler:2013xha}, where the infrared behaviour was regulated by introducing an effective hadronic mass scale. Since the value of such a scale is not fixed by the perturbative calculation, this prescription introduces an additional ambiguity in the resulting box correction. Our finite-mass calculation instead determines the low-energy result directly from the amplitude, without the need for an auxiliary infrared regulator.

For the phenomenological applications considered here, we therefore take the $m_q\to0$ limit. In deep-inelastic scattering, logarithmic quark-mass dependence can be separated from the perturbative hard contribution through the universal parton distribution functions. An explicit quark-mass logarithm in the box correction, however, carries coefficients depending on the electroweak couplings of the electron and quark and therefore cannot be absorbed into universal PDFs. Setting $m_q=0$ thus provides a perturbative contribution free from an arbitrary low-energy quark-mass scale, while allowing the remaining target-structure dependence to be treated separately.


\section{Reanalysis of PVDIS theoretical predictions with the $\gamma Z$-box correction}
\label{sec:reanalysis}

We now revisit the phenomenological analysis of the Jefferson Lab electron-deuteron measurement~\cite{Wang:2014guo}. The original analysis uses the Standard Model expressions for the effective electron-quark couplings given in Ref.~\cite{Erler:2013xha}, together with the input parameters available at the time. Here we update these parameters using the values adopted by the Particle Data Group~\cite{ParticleDataGroup:2024cfk}. Table~\ref{tab:ComparisonCiq} shows the extracted couplings before and after replacing the $\gamma Z$-box contribution by the well-defined finite-mass result presented in Sec.~\ref{sec:gammaZbox}. We retain the treatment of all other corrections as in Ref.~\cite{Wang:2014guo}, allowing us to isolate the impact of the modified $\gamma Z$-box contribution.

\begin{table}[]
\caption{Extracted effective electron-quark couplings. The first row collects the results of Ref.~\cite{Wang:2014guo}. The second row corresponds to the same analysis recalculated with the updated input parameters used here, while the third row additionally uses the $\gamma Z$-box contribution obtained in Sec.~\ref{sec:gammaZbox}.}

\centering
\label{tab:ComparisonCiq}
\begin{tabular}{ccccc}
\hline
           & $C_{1u}$ & $C_{1d}$ & $C_{2u}$ & $C_{2d}$ \\ \hline
Ref.~\cite{Wang:2014guo}                      & -0.1887 & 0.3419 & -0.0351 & 0.0248 \\
Ref.~\cite{Wang:2014guo} {[}updated inputs{]} & -0.1904 & 0.3419 & -0.0358 & 0.0257 \\
Our result & -0.1873  & 0.3454   & -0.0329  & 0.0290   \\ \hline
\end{tabular}
\end{table}

The aforementioned replacement of the $\gamma Z$-box  shifts all four extracted couplings towards larger values. To isolate this effect, we define $\Delta C_{iq}\equiv C_{iq}^{\rm this\ work}-C_{iq}^{\rm updated}$, where $C_{iq}^{\rm updated}$ denotes the second row of Table~\ref{tab:ComparisonCiq}. We find
\begin{equation}
\Delta C_{1u}=+0.0032,\qquad
\Delta C_{1d}=+0.0036,\qquad
\Delta C_{2u}=+0.0029,\qquad
\Delta C_{2d}=+0.0033.
\end{equation}
Although the shifts are small in absolute terms, they are relevant for a precision interpretation of the PVDIS measurement, particularly for the $C_{2q}$ couplings, to which PVDIS provides unique sensitivity.

A particularly relevant combination for PVDIS is $2C_{2u}-C_{2d}$. At tree level, the two $C_{2q}$ couplings are proportional to $1-4\sin^2\theta_W$ and therefore exhibit the same accidental suppression that makes the proton weak charge particularly sensitive to radiative corrections. Using the $\gamma Z$-box contribution adopted in Ref.~\cite{Wang:2014guo}, but with the updated electroweak inputs used here, the values of $C_{2u}$ and $C_{2d}$ in the second row of Table~\ref{tab:ComparisonCiq}, give $2C_{2u}-C_{2d}=-0.0973$. With the $\gamma Z$-box contribution calculated in Sec.~\ref{sec:gammaZbox}, this combination becomes $2C_{2u}-C_{2d}=-0.0948$, corresponding to a shift of $+0.0025$. The low-energy Standard Model value used in Ref.~\cite{Wang:2014guo}, $2C_{2u}-C_{2d}=-0.0950$, is obtained from the first row of Table~\ref{tab:ComparisonCiq}. The corresponding shift in $2C_{1u}-C_{1d}$ is $+0.0027$, showing that the modification affects both parity-violating coupling combinations.

The agreement of the original extraction with the low-energy Standard Model expectation at the time~\cite{Wang:2014guo} does not, by itself, establish the adequacy of the underlying box prescription. Our result shows that modifying this single contribution can shift the extracted $C_{1q}$ and $C_{2q}$ couplings at a level relevant to the precision interpretation of the measurement. A consistent treatment of the box contribution is therefore important when interpreting PVDIS measurements and deriving phenomenological constraints on the electron-quark couplings.

\section{Perturbative and Hadronic Contributions}
\label{sec:hadronic}
The perturbative result obtained from the electron-quark calculation is not, by itself, the complete $\gamma Z$-box correction for a hadronic target. Rather, it defines the perturbative contribution that can be matched onto the low-energy effective couplings, while the remaining contribution contains the non-perturbative dynamics of the target. In this sense, taking the $m_q\to0$ limit provides a well-defined perturbative separation without introducing an effective constituent-quark mass as an additional infrared scale.

This distinction is relevant not only for PVDIS but also for other precision parity-violation observables. In atomic parity violation, for example, the $\gamma Z$ box contributes to the nuclear weak charge. In Ref.~\cite{Duch:2026ujd}, the impact of the perturbative prescription was illustrated for $^{133}$Cs by comparing the $m_q\to0$ result obtained here with the conventional treatment using the $\rho$ mass as an infrared regulator, as in Ref.~\cite{Erler:2003yk}. The two prescriptions lead to a shift of about $-0.43$ in the extracted nuclear weak charge, corresponding to $\Delta\sin^2\theta_W\simeq0.0020$, comparable to the uncertainty quoted in Ref.~\cite{Sahoo:2021thl}.

This example illustrates the relevance of a well-defined perturbative contribution for precision parity-violation observables. A complete treatment requires the corresponding process-dependent non-perturbative contribution, whose consistent determination is an important direction for future work and for which our calculation provides a well-defined perturbative starting point without introducing an arbitrary effective quark-mass scale.

\section{Conclusions}

We have revisited the low-energy $\gamma Z$-box contribution using a finite-mass calculation, avoiding the introduction of an arbitrary effective quark-mass regulator. Applying this result to the Jefferson Lab PVDIS measurement leads to shifts of order $10^{-3}$ in the extracted electron-quark couplings, particularly relevant for the $C_{2q}$ combinations. Our results highlight the importance of a well-defined perturbative $\gamma Z$-box contribution for precision parity-violation analyses and provide a consistent starting point for incorporating the remaining non-perturbative hadronic effects.

\section*{Acknowledgment}
This work was supported by the DFG Cluster of Excellence PRISMA++ (EXC 2118/2, Project ID 390831469), and by the Ministerio de Ciencia e Innovación (PID2023-146142NB-I00), and through the Severo Ochoa Programme (CEX2024-001441-S) funded by MICIU/AEI/10.13039/ 501100011033. This project has also received funding from the European Research Council (ERC) under the
European Union’s Horizon research and innovation programme (Grant agreement No. 101142600). IFAE is partially funded by CERCA. B.~D.\ also acknowledges support from the AGAUR-FI Joan Oró programme (2025 FI-1 00461), co-financed by the European Social Fund Plus.


\end{document}